\documentclass[10pt,twocolumn,letterpaper]{article}

\usepackage{iccv}
\usepackage{times}
\usepackage{epsfig}
\usepackage{graphicx}
\usepackage{amsmath}
\usepackage{amssymb}
\usepackage{float}
\usepackage{subcaption}
\usepackage{algorithm}
\usepackage{algorithmic}
\usepackage{gensymb}
\usepackage{booktabs}
\usepackage{multirow}
\usepackage{array}
\usepackage{makecell}
\usepackage{setspace}
\usepackage{pifont}
\usepackage{float}
\usepackage{colortbl}
\usepackage{arydshln}

\usepackage[pagebackref=true,breaklinks=true,letterpaper=true,colorlinks,bookmarks=false]{hyperref}

\iccvfinalcopy 

\def\iccvPaperID{7181} 

\ificcvfinal\fi

\begin{document}

\title{Learning Frequency-aware Dynamic Network for Efficient Super-Resolution}

\author{Wenbin Xie$^{1,2*}$, Dehua Song$^{1} $\thanks{Equal contribution} , Chang Xu$^{3}$, Chunjing Xu$^{1}$, Hui Zhang$^{2}$, Yunhe Wang$^{1}$ \\
\normalsize$^1$ Noah's Ark Lab, Huawei Technologies.  \ \ \normalsize$^2$ School of Software, Tsinghua University. \ \ \normalsize$^3$ The University of Sydney.\\
\small\texttt{\{dehua.song, yunhe.wang, xuchunjing\}@huawei.com;} \small\texttt{xiewb18@mail.tsinghua.edu.cn;}
}

\maketitle
\ificcvfinal\thispagestyle{empty}\fi

\begin{abstract}

Deep learning based methods, especially convolutional neural networks (CNNs) have been successfully applied in the field of single image super-resolution (SISR). To obtain better fidelity and visual quality, most of existing networks are of heavy design with massive computation. However, the computation resources of modern mobile devices are limited, which cannot easily support the expensive cost. To this end, this paper explores a novel frequency-aware dynamic network for dividing the input into multiple parts according to its coefficients in the discrete cosine transform (DCT) domain. In practice, the high-frequency part will be processed using expensive operations and the lower-frequency part is assigned with cheap operations to relieve the computation burden. Since pixels or image patches belong to low-frequency areas contain relatively few textural details, this dynamic network will not affect the quality of resulting super-resolution images. In addition, we embed predictors into the proposed dynamic network to end-to-end fine-tune the handcrafted frequency-aware masks. Extensive experiments conducted on benchmark SISR models and datasets show that the frequency-aware dynamic network can be employed for various SISR neural architectures to obtain the better tradeoff between visual quality and computational complexity. For instance, we can reduce the FLOPs of SR models by approximate $50\%$ while preserving state-of-the-art SISR performance.

\end{abstract}

\section{Introduction}

\begin{figure}[t]
	\begin{center}
		\includegraphics[width=1\columnwidth]{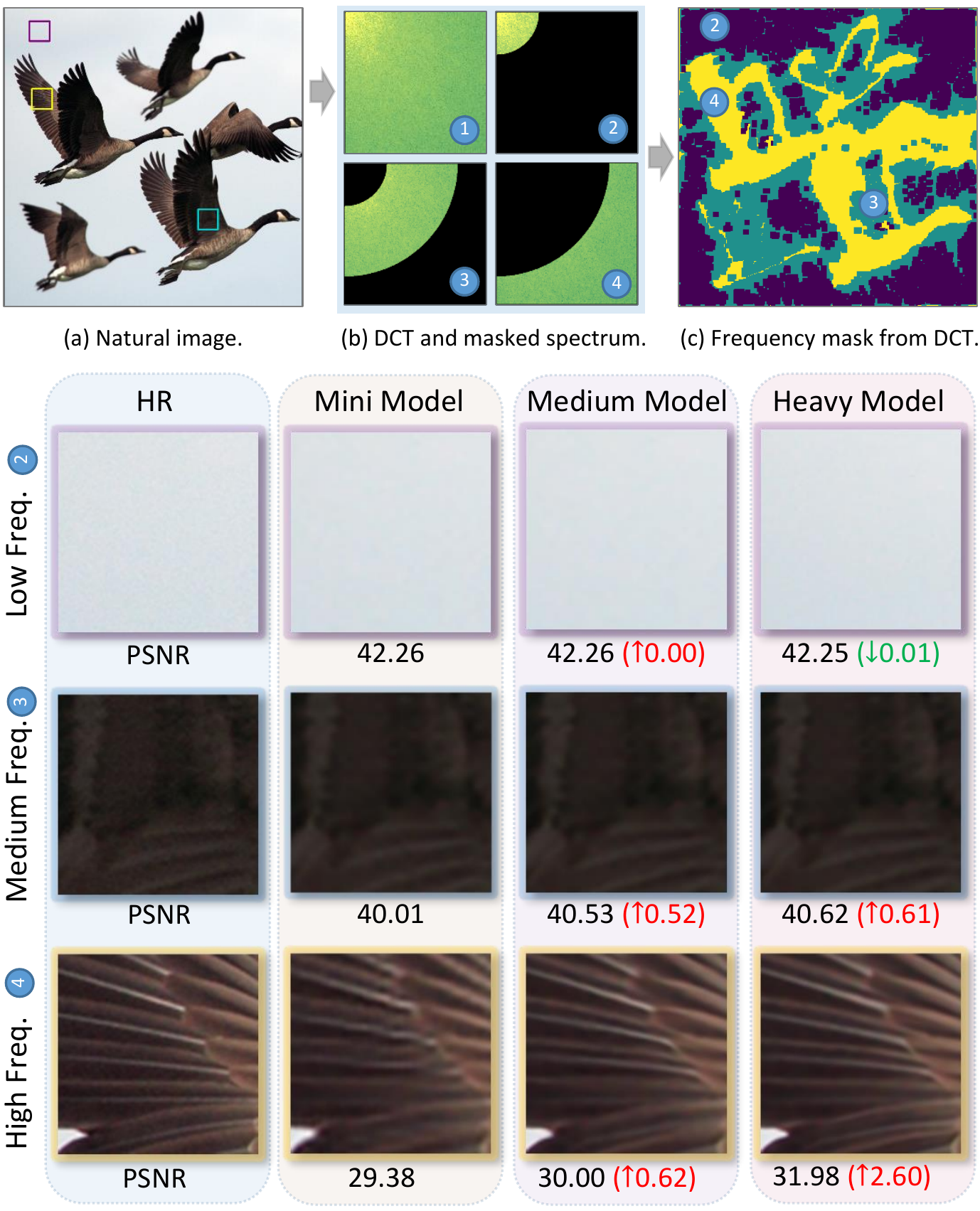}
	\end{center}
\vspace{-0.5cm}
\caption{The motivation of our frequency-aware dynamic network. We illustrate the performance of SISR networks with various amounts of computation separately processing patches with different frequency signals. The profit brought by heavy computations becomes slighter as the frequency decreases. For conventional SR models, massive computation redundancy exists in processing low and medium frequency regions. Wherein, \ding{173}, \ding{174} and \ding{175} denote low, medium and high frequency signals/regions, respectively. 
}
\label{fig:introduction}
\vspace{-0.5cm}
\end{figure}

Single image super resolution (SISR) receives low-resolution images and outputs their high-resolution counterparts, which is widely used in real-world applications such as mobile phone, surveillance, autonomous driving, \etc. Basically, SISR is an ill-posed reverse problem for recovering more information from the low-resolution versions. Thanks to the great progress of deep learning, a number of approaches have been explored using deep convolutional neural networks (CNNs) for addressing the SISR problem. Since neural networks can capture more information from a large amount of available images thus yield higher performance over conventional image recovery algorithms.

Similar to most computer vision tasks, the design of network architectures is quite important for the performance of SISR. Dong \etal~\cite{dong2014learning} first employed a network on super-resolution with only three convolutional layers, which obtained better performance than traditional methods. Subsequently, a series of networks with sophisticated architectures and loss functions are developed. For instance, Lim~\etal~\cite{lim2017enhanced} deepened the SR network with 32 resblocks. Tai~\etal~\cite{tai2017memnet} and Zhang~\etal~\cite{zhang2018residual} investigated the dense concatenation on SR. In addition, channel attention (\eg, RCAN~\cite{zhang2018image} and SAN~\cite{dai2019second}) and spatial attention (\eg, ABPN~\cite{liu2019image}) mechanisms were also embedded in SISR models and boosted the performance significantly.

Although tremendous efforts have been made to refine quantitative results \ie, PSNR (Peak Signal-to-Noise Ratio) and SSIM (Structure Similarity), and visual quality of generated high-resolution images, the computational cost should be carefully restrained for real-world applications. For instance, 10,194G FLOPs is required for generating a $1280 \times 720$ (720p) image using RDN~\cite{song2019efficient} model. To improve the model efficiency for deploying them on mobile devices while retaining the performance, Ahn~\etal~\cite{ahn2018fast} and Luo~\etal~\cite{luolatticenet} employed cheap operators to construct efficient SR models manually. Furthermore, Chu~\etal~\cite{chu2019fast} and Song~\etal~\cite{song2019efficient} explored neural architecture search (NAS) to acquire efficient SISR networks automatically.

Nevertheless, most of existing approaches focus on reducing computation by processing the whole image with the same way, which are not perfectly efficient. Natural images are composed of distinct frequency signals according to the Fourier Transform~\cite{bracewell1986fourier}. Recovering high frequency information requires massive computations due to its severe damage during the downsample procedure, but reconstructing low frequency information does not demand such huge computations. Fig.~\ref{fig:introduction} illustrates this phenomenon and indicates that the profit brought by heavy computations becomes slighter as the frequency decreases. Hence, massive computation redundancy exists in processing low and medium frequency regions. It motivates us to explore a more efficient SR method according to frequencies of the input instance.

To this end, this paper proposes a novel frequency-aware dynamic convolutional network (FADN). In each block, we introduce a predictor for dividing the input feature into multiple components based on the discrete cosine transform (DCT \cite{ahmed1974discrete}) domain, \eg, high-frequency, medium-frequency and low-frequency parts. The predictor is learned under the supervision of hand-crafted frequency-domain masks of images in the training set and the reconstruction loss of SISR, simultaneously. Then, features in these multiple parts will be processed using different convolutional layers with various computation burdens. The features with only low-frequency domain information will be assigned with cheaper operations for reducing computations and vice versa. Since features are divided into multiple branches in our paradigm, the overall computational complexity will be significantly reduced by the optimized allocation.

Extensive experiments are carefully conducted to verity the effectiveness of the proposed frequency-aware dynamic network on mainstream super-resolution benchmarks. The experimental results demonstrate that our method is able to employ on different neural architectures for achieving comparable and even better super-resolution performance using fewer computations.

The rest of this paper is structured as follows: we firstly summarize the related works on super-resolution methods in Section~\ref{RelatedWorks}. In Section \ref{method}, the proposed method is introduced in detail. Then, comparison experiments and ablation study are depicted in Section \ref{experiment}. At last, we draw conclusions of the paper in Section \ref{conclusion}.

\begin{figure*}[!h]
    \begin{center}
        \includegraphics[width=1\textwidth]{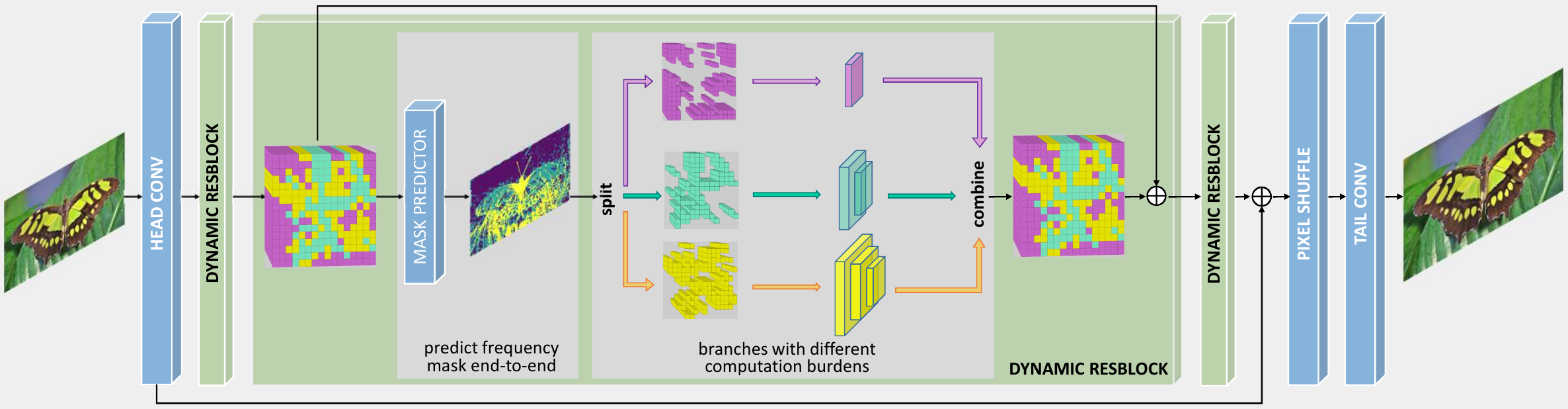}
    \end{center}
\vspace{-0.4cm}
\caption{The overall framework of the proposed frequency-aware dynamic SR network (FADN).}
\label{fig:framework}
\vspace{-0.2cm}
\end{figure*}

\section{Related Work}
\label{RelatedWorks}
SISR problem has been investigated broadly in the last decades and remarkable improvement has been acquired. We summarize and analyze various SR methods.

\subsection{Superior Super-resolution Methods}

Enormous super-resolution methods can be roughly divided into three categories: interpolation-based, reconstruction-based and learning-based methods. Interpolation methods are efficient but suffer from severe flaws of fidelity \cite{keys1981cubic}. In order to generate flexible and sharp details, reconstruction-based methods exploit sophisticated prior knowledge to reconstruct the high-resolution image \cite{yan2015single}. However, these methods are time-consuming and sensitive to the handcrafted parameters. Recently, deep learning methods have dramatically boosted the performance of SISR. Dong \etal~\cite{dong2014learning} firstly introduced DCNNs to SISR with only three convolution layers, yet its performance was greatly superior to that of traditional methods. Then, many approaches exploited the depth of DCNNs to improve the fidelity of SISR further, such as VDSR~\cite{kim2016accurate}, SRResNet~\cite{ledig2017photo}, and EDSR~\cite{lim2017enhanced}. In addition, MemNet~\cite{tai2017memnet} and RDN~\cite{zhang2018residual} also explored dense concatenation to fusion features with different receptive fields. Channel attention (\eg RCAN~\cite{zhang2018image}, SAN~\cite{dai2019second}) and spatial attention (\eg ABPN~\cite{liu2019image}) were both applied on super-resolution network for better recovery of details. Lately, non-local-based graph networks (\eg GCDN~\cite{valsesia2020deep}, IGNN~\cite{zhou2020cross}) are becoming a new trend to recover the detailed textures. On the other hand, perceptual loss~\cite{johnson2016perceptual} and adversarial training strategy~\cite{ledig2017photo} were adopted to improve the visual quality.  

\subsection{Efficient Super-resolution Models}
The fidelity of SISR has been promoted significantly while the computational cost increases rapidly too. Massive computations of networks severely limited its application on real-world mobile devices. Hence, many researchers shifted their attention to developing efficient SISR networks. Dong \etal~\cite{dong2016accelerating} firstly accelerated SISR models by delaying the upsample operation. Then, researchers began to design efficient blocks or networks with cheap operators, for example, CARN~\cite{ahn2018fast}, IDN~\cite{hui2018fast,hui2019lightweight} and LatticeNet~\cite{luolatticenet}. To exploit efficient architecture thoroughly, neural architecture search (NAS) methods (\eg FALSR~\cite{chu2019fast}, ESRN~\cite{song2019efficient}) were employed to seek for extreme lightweight SISR networks. Besides, knowledge distillation technique~\cite{gao2018image,yang2020distilling} was introduced to SISR to take full advantage of teacher network. Lately, quantization SISR networks~\cite{ma2019efficient,xin2020binarized} were also explored with the assistance of feature complementary.

Besides, Liu \etal \cite{liu2020deep} designed a dynamic inference network to reduce computations, by predicting a depth mask and ignoring the calculations of partial features once the layer is deeper than the predicted depth. Note that the depth predictor cannot be optimized end-to-end. Verelst \etal \cite{verelst2020dynamic} proposed a similar dynamic network that elides partial calculations for the classification task. These dynamic convolutions are totally designed on the spatial domain. Instead, we proposal a novel dynamic mechanism based on the frequency-domain information in this paper. Our method recovers distinct frequency signals with branches requiring different computations and thus reduces the computational complexity.

\section{Frequency-aware Dynamic Network}
\label{method}
The framework of our proposed frequency-aware dynamic network is shown in Fig.~\ref{fig:framework}. FADN consists of feature extraction, frequency-aware dynamic blocks (FADB) and reconstruction. Each dynamic block contains multiple branches with different computation burdens. In the inference stage, FADB automatically assigns higher frequency regions to heavier branches and lower frequency regions to lighter branches according to the learnable frequency mask.

\subsection{Frequency-aware Dynamic Block}
\begin{figure}[t]
    \begin{center}
    \begin{subfigure}[t]{0.36\columnwidth}
        \includegraphics[width=1\textwidth]{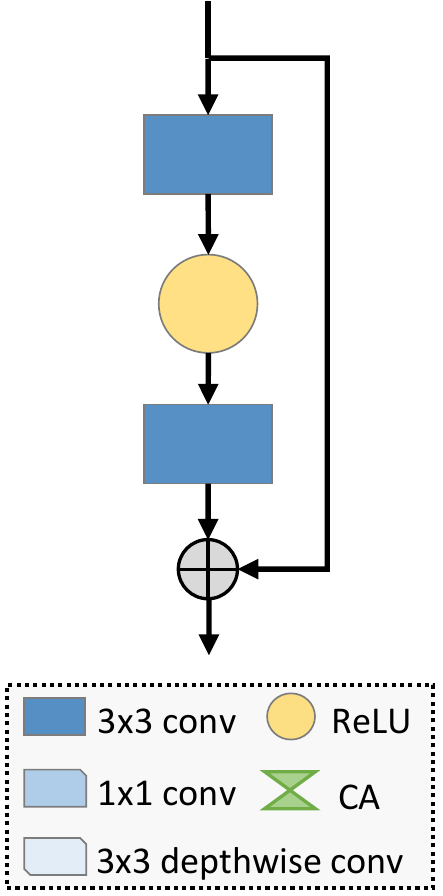}
        \caption{}
    \end{subfigure}
    \begin{subfigure}[t]{0.21\columnwidth}
        \includegraphics[width=1\textwidth]{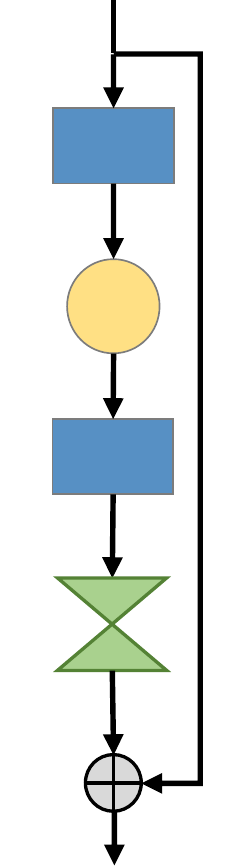}
        \caption{}
    \end{subfigure}
    \begin{subfigure}[t]{0.405\columnwidth}
        \includegraphics[width=1\textwidth]{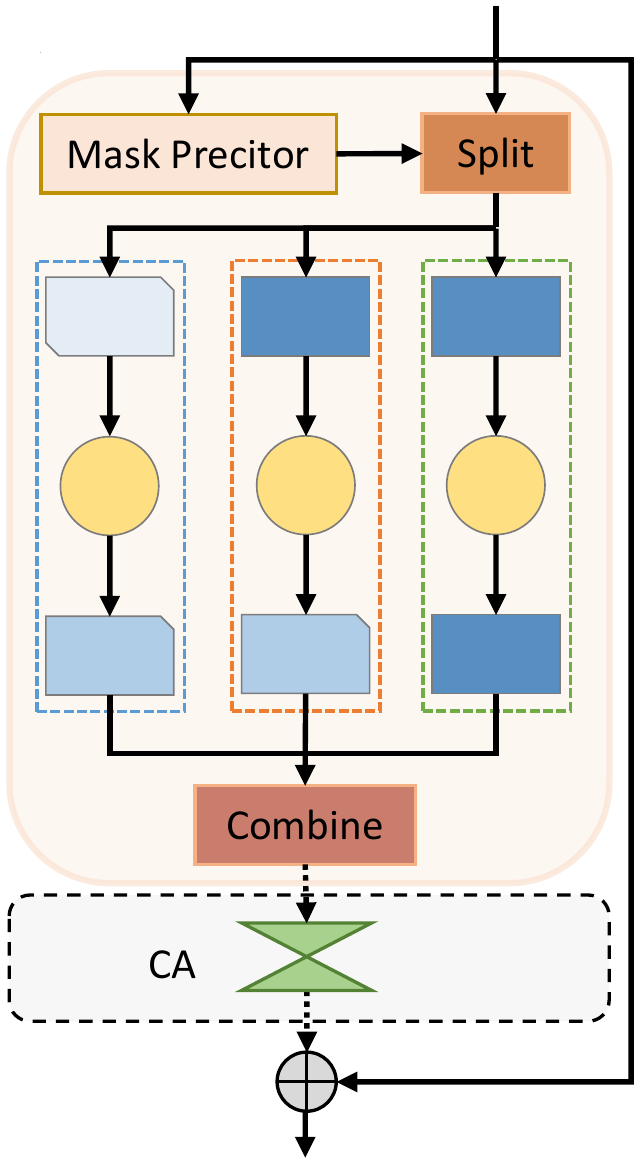}
        \caption{}
        \label{fig:dynamicresblock}
    \end{subfigure}
    \end{center}
\vspace{-0.4cm}
       \caption{The structure of prominent SR blocks. (a) Block in EDSR~\cite{lim2017enhanced}. (b) Block in RCAN~\cite{zhang2018image} (CA operator stands for channel attention). (c) Corresponding frequency-aware dynamic block (CA is only employed in FAD-RCAN).}
    \label{fig:resblock}
    \vspace{-0.3cm}
\end{figure}

Firstly, we revisit prominent blocks in SR task, for example, resblock in EDSR \cite{lim2017enhanced} and channel attention resblock in RCAN~\cite{zhang2018image}. The detail architectures of blocks are shown in Fig.~\ref{fig:resblock}. Block contains two $3 \times 3$ convolution and one ReLU layers. In addition, RCAB also include a channel attention operator. Notating the input tensor as $X$ and the output tensor as $Y$, RCAB can be modeled as
\begin{equation}
    Y = X + CA(F),
\end{equation}
where $CA$ denotes the channel attention operator, $F$ is output feature of the first two convolution layers, $F = f(X)$.

Channel attention averages the features across spatial space, hence our frequency-aware dynamic mechanism focuses on the feature transformation stage $f$. In the standard SR block, all pixels share same filters (\ie filter size, number and weights). However, the pixels should not be treated equally in SR task, since high frequency signals suffer more seriously damage than low frequency signals during downscaling. In other words, low frequency signals can be recovered using cheap operations. Based on the idea, we design the frequency-aware dynamic block (FADB) which contains multiple branches with different burdens. In the inference stage, each pixel is assigned to a specific branch according to a learnable frequency-aware mask $M$. The number and detail structures of branches are optional. A typical FADB setting is shown in Fig.~\ref{fig:dynamicresblock} which includes three branches. The heaviest branch is the same with standard block. The computations decrease gradually from left two branches to the right one. Notating the branch number as $K$ and the function of branch $k$ as $f_k$, the FADB can be represented as

\begin{equation}
\begin{aligned}
    \label{equation:dynamic_block}
    & F_i = \sum_{k=1}^K M_{i, k} \cdot f_k(X_i), \\
    & s.t. \ \ M_{i, k} \in \{0, 1\} \ \text{\it and} \ \sum_k^K M_{i, k} = 1.
 \end{aligned}
\end{equation}
where $i$ is the pixel index. The constrains mean that only one branch $k$ where $M_{i, k} = 1$ will be chosen for $X_i$.

\subsection{Mask Predictor}
\label{section:maskpredictor}
It's important to assign $X_i$ to a proper branch according to the frequency strength so that lower frequency signals can be recovered using lighter branches to save computations.
An intuitive strategy is to apply DCT type \uppercase\expandafter{\romannumeral2}~\cite{ahmed1974discrete, strang1999discrete} to generate a frequency mask. We first use DCT to convert the image from spatial domain to frequency domain as shown in Fig.~\ref{fig:introduction}~(b-1) where the upper left parts are for low frequency signals and lower right parts are for high frequency signals. Then, we employ $T$ frequency thresholds to separate the spectrum into $T + 1$ parts (\eg, 2 thresholds generate 3 spectrum parts as shown in Fig.~\ref{fig:introduction}~(b-2,3,4)). At last, these $T + 1$ spectrum parts are reconverted to spatial domain via IDCT, respectively, and generate the DCT frequency mask as shown in Fig.~\ref{fig:introduction}~(c). During the reconverting stage, we process high frequency preferentially, \ie, once a pixel belongs to higher frequency, it won't be considered as lower frequency in the later processing. Hence, we usually generate DCT frequency mask from high frequency region to low frequency region sequentially.

Unfortunately, human-crafted thresholds are sensitive to noise and specific values are not suitable for all images. We expect the network can learn a frequency mask indicating which branch will be adopted by each pixel in an end-to-end manner. We make use of Gumbel Softmax trick~\cite{jang2016categorical} to make the mask prediction differentiable.

Given $X$, a $1 \times 1$ convolution is used as a simple mask predictor to generate a distribution $D$ of $K$ channels and same spatial size as $X$. The predictor is so small that its  computational cost can be almost ignored. The frequency mask $M$ can be calculated as

\begin{equation}
	\label{equation:naive_forward}
	M_i = \mathop{one\_hot}(\mathop{\arg\max}_k \ D_{i, k}).
\end{equation}
Equation~\ref{equation:naive_forward} has two drawbacks. First, it ignores the statistic randomness as $D$ is actually a continuous distribution instead of discrete variable. Second, it's not differentiable and hence can not be optimized end-to-end. Therefore, we apply Gumbel Softmax trick to get a continuous, differentiable normalized distribution:
\begin{equation}
    \label{equation:gumbel}
    G_i= \frac{exp( (D_i + g_i) / \tau )} { \sum_{k=1}^K{exp((D_{i, k} + g_{i, k})/\tau)} },
\end{equation}
where $g_i$ is drawn from $Gumbel(0, 1)$, $\tau$ is a temperature value controlling the distribution density and set as $1$ in our experiments. The noise is only added in the training stage. We hope only one branch will be chosen for each pixel. In the forward process, the chosen branch (\ie the frequency mask) is
\begin{equation}
    \label{equation:forward}
    M_i = \mathop{one\_hot}(\mathop{\arg\max}_k \ G_{i, k}).
\end{equation}
In order to train the predictor end-to-end, we calculate the gradient of $M$ according to Equation~\ref{equation:gumbel} in the backward process, \ie,
\begin{equation}
	M_i = \begin{cases}
	\mathop{one\_hot}(\mathop{\arg\max}\limits_{k} \ G_{i, k}) &forward, \\
	G_i &backward.
	\end{cases}
\end{equation}

\subsection{Loss}
The proposed FADN is frequency-aware that models low frequency signals with cheaper operations. To this end, the mask predictor should absorb the prior knowledge from DCT frequency domain. Specifically, we make use of the frequency mask generated with DCT, notated as $D$, to guide the mask training. Assuming there are $B$ dynamic resblocks, the DCT loss is 
\begin{equation}
    L_{dct} = \sum_{b=1}^B \text{\it CE}(M_b, \mathop{D}),
\end{equation} where $\text{\it CE}(\cdot)$ is cross entropy loss and $M_b$ is the predicted frequency mask of the $b$-th block.

Besides, it's important to control the computational cost conveniently for satisfying the requirements of various scenes.  We realize that the sparsity of predicted frequency mask reflects the total computations, \ie, when the predicted frequency mask contains more low frequency components, the computational cost is smaller. Assuming there are $n$ pixels in each block and $c_k$ represents per-pixel FLOPs of branch $k$, the total FLOPs of a dynamic block are
\begin{equation}
    \label{equation:flops_dynamic_block}
    C = \sum_{i=1}^n \sum_{k=1}^K M_{i, k} \cdot c_k.
\end{equation} 
The sparsity loss is designed as
\begin{equation}
	\label{equa:sparsity_loss}
    L_{spa} = ( \frac{ \sum_{b=1}^B C_b }{B \cdot n c_K} - \alpha )^2,
\end{equation}
where $C_b$ is FLOPs of the $b$-th block, $n c_K$ is FLOPs of a SR block and $\alpha$ is a hyper-parameter to control the frequency mask sparsity as well as the total FLOPs.

The total frequency mask loss is
\begin{equation}
    L_{mask} = L_{spa} + \beta L_{dct},
\end{equation}
where $\beta$ is a hyper-parameter to balance the two mask losses. We apply an annealing strategy on $\beta$ and reduce it to zero gradually during training considering the following two reasons. First, the DCT frequency mask is not accurate since it is dependent on human-crafted thresholds and sensitive to noises. Second, $L_{dct}$ cannot control the expected computations conveniently. Therefore, only $L_{spa}$ is used to guide the mask learning in the later stage for better convergence and more precise control of computations.

To make the recovered images are of similar visual quality as the origin high resolution versions, we use L1 distance as the super-resolution loss:
\begin{equation}
    L_{sr} = \|sr - hr\|_1,
\end{equation}
where $sr$ is the super-resolution image and $hr$ is the ground truth as well as high-resolution image.

Finally, FADN is optimized by the two kinds of losses, $L_{sr}$ for recovering more details and $L_{spa}$ for guiding the predictor to learn the frequency distribution and control computational cost. The complete loss can be defined as
\begin{equation}
    L = L_{mask} + L_{sr}.
\end{equation}

\subsection{Efficient Implementation}
Since weight sharing is still kept along each branch, the calculation can be still implemented by GEMM \cite{lawson1979basic} effectively.
The first convolution in each branch should have same kernel size (\eg 3) so that we can first apply {\it image2col} to unfold the input $X$. Meanwhile, the mask predictor generates frequency mask $M$. Then, the unfolded $X$ is split and serves as the input of each branch according to $M$. The first convolution can be taken as the matrix multiplication between the unfolded input and kernel weight. $1 \times 1$ convolution and ReLU module is easy to implement since they are pixel-independent. The corresponding frequency mask should be dilated by a $3 \times 3$ kernel to keep receptive field if the branch includes two $3 \times 3$ convolution layers. After calculating the results of each branch, we combine them into an entire tensor.

\subsection{Discussion}
\textbf{FADN \vs other spatially-varying computing methods} Verelst \etal \cite{verelst2020dynamic} proposed a dynamic mechanism by exploiting spatial sparsity for classification. It pays more computation on semantic contents and is not suitable for SISR task. AdaDSR~\cite{liu2020deep} predicts the depth for each pixel directly. All these methods developed a sparse convolution that neglects the calculations of partial input features once meeting certain conditions, which causes partial inputs cannot be modeled effectively. In contrast, based on the characteristics of SR illustrated in Fig.~\ref{fig:introduction}, our method takes full use of frequency-domain information to construct dynamic block since frequency has a positive relation with recovery difficulty of SR. Hence, the proposed FADN is quite different from other spatially-varying computing methods.

\textbf{FADN \vs OctaveConv.} The Octave convolution~\cite{chen2019drop} separates the high-frequency and low-frequency with down-sample operations. In essence, the main idea of Octave convolution is consistent with multi-resolution feature representation. In contrast, the proposed dynamic network separates the regions with different frequency information according to mask predictor, which is consistent with DCT frequency information. Their main ideas are quite different.

\begin{table*}[ht]
	\renewcommand\arraystretch{1.2}
	\small 
	\begin{center}
		\caption{Quantitative results in comparison with the state-of-the-art methods on four benchmark databases. The best results are highlighted in bold.}
		\resizebox{\textwidth}{!}{
			\begin{tabular}[]{|c|l|c|c|c|c|c|c|c|c|c|c|c|c|}
				\hline
				\multirow{2}{*} {Scale} & \multirow{2}{*}{Method} & \multicolumn{3}{c}{Set5} & \multicolumn{3}{|c}{Set14} & \multicolumn{3}{|c}{B100} & \multicolumn{3}{|c|}{Urban100} \\
				\cline{3-14}
				& & \makecell[c]{PSNR $\uparrow$ \\ (dB)}  & SSIM  $\uparrow$ & \makecell[c]{FLOPs $\downarrow$ \\ (G)} & \makecell[c]{PSNR  $\uparrow$ \\ (dB)} & SSIM  $\uparrow$ & \makecell[c]{FLOPs $\downarrow$ \\ (G)} & \makecell[c]{PSNR  $\uparrow$ \\ (dB)} & SSIM $\uparrow$ & \makecell[c]{FLOPs $\downarrow$ \\ (G)} & \makecell[c]{PSNR  $\uparrow$ \\ (dB)} & SSIM $\uparrow$ & \makecell[c]{FLOPs $\downarrow$ \\ (G)} \\
				\hline
				
				\multirow{8}{*}{$\times 2$} & Bicubic & 33.66 & 0.9299 &  --- & 30.24 & 0.8688 & --- & 29.56 & 0.8431 & --- & 26.88 & 0.8403 & --- \\
				& VDSR~\cite{kim2016accurate} & 37.53 & 0.9590 & 70.5  & 33.05 & 0.9130 & 143.0  & 31.90 & 0.8960 & 95.4  & 30.77 & 0.9140 & 481.6 \\ 
				\cdashline{2-14}[3pt/2pt]
				& EDSR~\cite{lim2017enhanced} & 38.11 & 0.9601 & 1338.8 & 33.92 & 0.9195 &  2552.2 & 32.32 & 0.9013 &  1776.9 & 32.93 & 0.9351 & 8041.1 \\ 
				& AdaEDSR~\cite{liu2020deep}  & 38.21 & 0.9611 & 650.6  & {\bf 33.97} & {\bf 0.9208} & 1397.3  & {\bf 32.35} & {\bf 0.9017} & 965.3  & 32.91 & 0.9353 & 4844.9  \\ 
				& {\bf FAD-EDSR}              & {\bf 38.21} & {\bf 0.9611} & {\bf 408.4}  & 33.95 & 0.9202 & {\bf 1068.8}  & 32.33 & 0.9015 & {\bf 686.7}  & {\bf 32.93} & {\bf 0.9353} & {\bf 4192.9}  \\ 
				\cdashline{2-14}[3pt/2pt]
				& RCAN~\cite{zhang2018image}  & 38.27 & 0.9614 &  577.9  & 34.12 & 0.9216 &  1101.8 & 32.41 & 0.9027 & 767.0  & 33.34 & 0.9384 & 3471.2\\ 
				& AdaRCAN~\cite{liu2020deep}  & 38.28 & 0.9615 & 469.1  & \bf{34.12} & \bf{0.9216} & 751.9  & 32.41 & 0.9026 & 606.3  & 33.29 & 0.9380 & 2907.2  \\ 
				& {\bf FAD-RCAN}              & {\bf 38.29} & {\bf 0.9617} & {\bf 260.8}  & 34.11 & 0.9215 & {\bf 512.6}  & {\bf 32.42} & {\bf 0.9028} & {\bf 363.7}  & {\bf 33.34} & {\bf 0.9385} & {\bf 2089.4}  \\ 
				\hline

				\multirow{8}{*}{$\times 3$} & Bicubic   & 30.39 & 0.8682 & --- & 27.55 & 0.7742 & ---  & 27.21 & 0.7385 & --- & 24.46 & 0.7349 & --- \\ 
				& VDSR~\cite{kim2016accurate} & 33.67 & 0.9210 &  70.5  & 29.78 & 0.8320 & 143.0  & 28.83 & 0.7990 & 95.4  & 27.14 & 0.8290 & 481.6 \\ 
				\cdashline{2-14}[3pt/2pt]
				& EDSR~\cite{lim2017enhanced} & 34.65 & 0.9280 &  699.1  & 30.52 & 0.8462 &  1305.7 & 29.25 & 0.8093 &  924.1 & 28.80 & 0.8653 & 3984.0 \\ 
				& AdaEDSR~\cite{liu2020deep}  & 34.65 & 0.9288 &  504.8  & 30.57 & 0.8463 &  1013.5  & 29.27 & 0.8091 & 722.8 & 28.78 & 0.8649 & 3314.2\\ 
				& {\bf FAD-EDSR}                  & {\bf 34.69} & {\bf 0.9288}  & {\bf 281.0} & {\bf 30.58} & {\bf 0.8467} & {\bf 651.5} & {\bf 29.27} & {\bf 0.8097} & {\bf 423.6} & {\bf 28.89} & {\bf 0.8668} & {\bf 2472.9} \\ 
				\cdashline{2-14}[3pt/2pt]
				&  RCAN~\cite{zhang2018image}  & 34.74 & 0.9299 & 328.5 & 30.65 & 0.8482 & 613.5 & 29.32 & 0.8111 & 434.2 & 29.09 & 0.8702 & 1872.0 \\ 
				& AdaRCAN~\cite{liu2020deep}  & 34.79 & 0.9302 & 277.7  & 30.65 & 0.8481 & 512.9  & 29.33 & 0.8111 & 369.3  & 29.03 & 0.8689 & 1596.3 \\ 
				& {\bf FAD-RCAN}              & {\bf 34.82} & {\bf 0.9304} & {\bf 141.2}  & {\bf 30.66} & {\bf 0.8485} & {\bf 306.1}  & {\bf 29.33} & {\bf 0.8112} & {\bf 212.3}  & {\bf 29.12} & {\bf 0.8706} & {\bf 1081.6}  \\ 
				\hline
				
				\multirow{8}{*}{$\times 4$} & Bicubic             & 28.42 & 0.8104 &  ---  & 26.00 & 0.7027 & --- & 25.96 & 0.6675 &  --- & 23.14 & 0.6577 &  --- \\ 
				& VDSR~\cite{kim2016accurate} & 31.35 & 0.8830 & 70.5 & 28.02 & 0.7680 & 143.0 & 27.29 & 0.0726 & 95.4 & 25.18 & 0.7540 & 481.6 \\
				\cdashline{2-14}[3pt/2pt]
				& EDSR~\cite{lim2017enhanced} & 32.46 & 0.8968 & 501.9 & 28.80 & 0.7876 & 908.8  & 27.71 & 0.7420 & 655.7 & 26.64 & 0.8033 & 2699.4 \\ 
				& AdaEDSR~\cite{liu2020deep}  & 32.49 & 0.8977 & 371.7 & 28.76 & 0.7865 & 716.8 & 27.71 & 0.7410 & 508.5 & 26.58 & 0.8011  & 2265.8 \\ 
				& {\bf FAD-EDSR}                  & {\bf 32.50} & {\bf 0.8977} & {\bf 218.1} & {\bf 28.82} & {\bf 0.7880} & {\bf 471.7} & {\bf 27.73} & {\bf 0.7438} & {\bf 304.6} & {\bf 26.70} & {\bf 0.8049} & {\bf 1729.9}  \\
				\cdashline{2-14}[3pt/2pt]
				&  RCAN~\cite{zhang2018image}  & 32.63 & 0.9002 & 270.1 & 28.87 & 0.7889 & 489.0 & 27.77 & 0.7436 & 352.8 & 26.82 & 0.8087 & 1452.5 \\ 
				& AdaRCAN~\cite{liu2020deep}  & 32.61 & 0.8998 & 277.5  & 28.88 & 0.7883 & 418.1 & 27.77 & 0.7428 & 304.8 & 26.80 & 0.8067 & 1263.0 \\ 
				& {\bf FAD-RCAN}              & {\bf 32.65} & {\bf 0.9007} & {\bf 115.4}  & {\bf 28.88} & {\bf 0.7889} & {\bf 258.3}  & {\bf 27.78} & {\bf 0.7437} & {\bf 169.8}  & {\bf 26.86} & {\bf 0.8092} & {\bf 891.2}  \\ 
				\hline
			\end{tabular}
		}
		\label{table:sota}
		\vspace{-0.3cm}
	\end{center}
\end{table*}

\begin{figure*}[ht]
	\begin{center}
		\begin{subfigure}[]{0.32\textwidth}
			\includegraphics[width=1\textwidth]{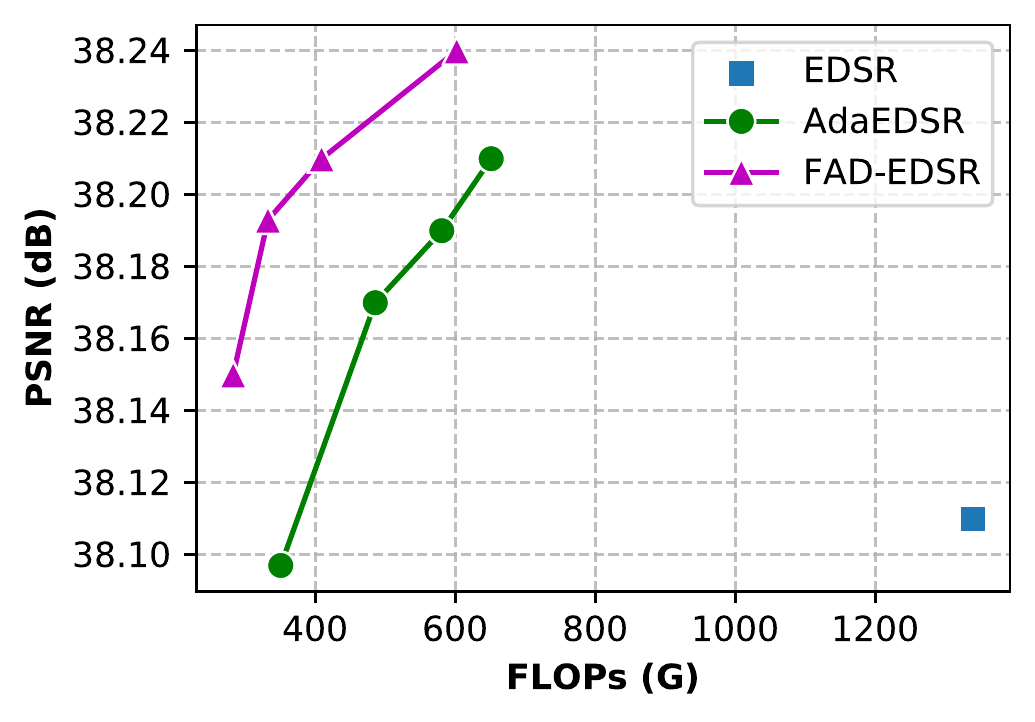}
			\vspace{-0.6cm}
			\caption{$\times 2$ scale.}
		\end{subfigure}
		\begin{subfigure}[]{0.32\textwidth}
			\includegraphics[width=1\textwidth]{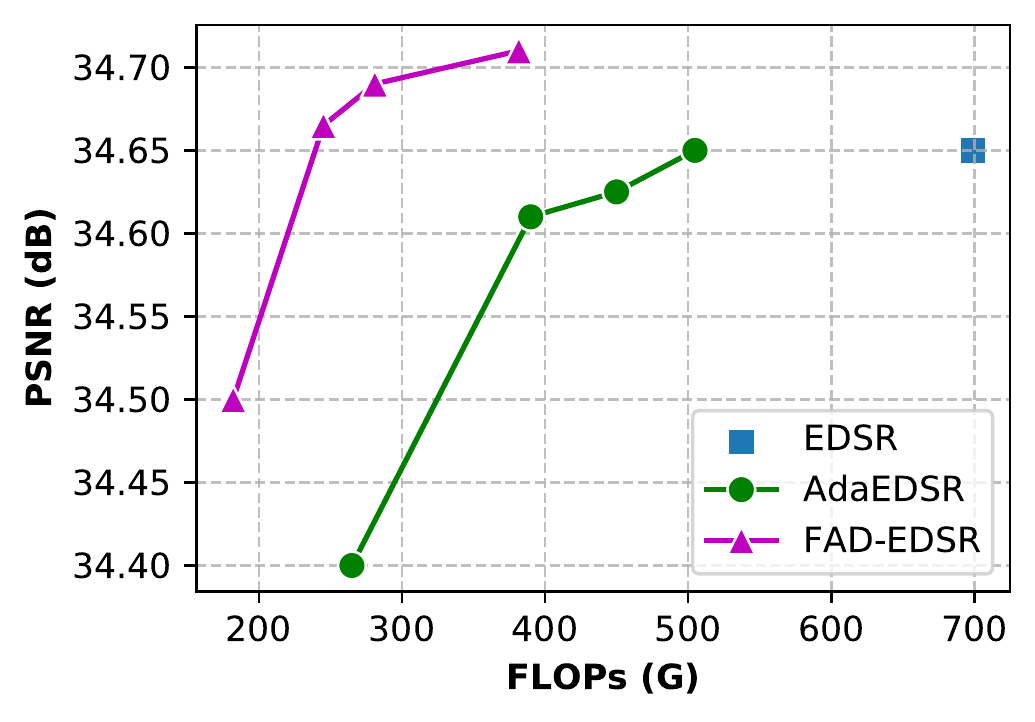}
			\vspace{-0.6cm}
			\caption{$\times 3$ scale.}
		\end{subfigure}
		\begin{subfigure}[]{0.32\textwidth}
			\includegraphics[width=1\textwidth]{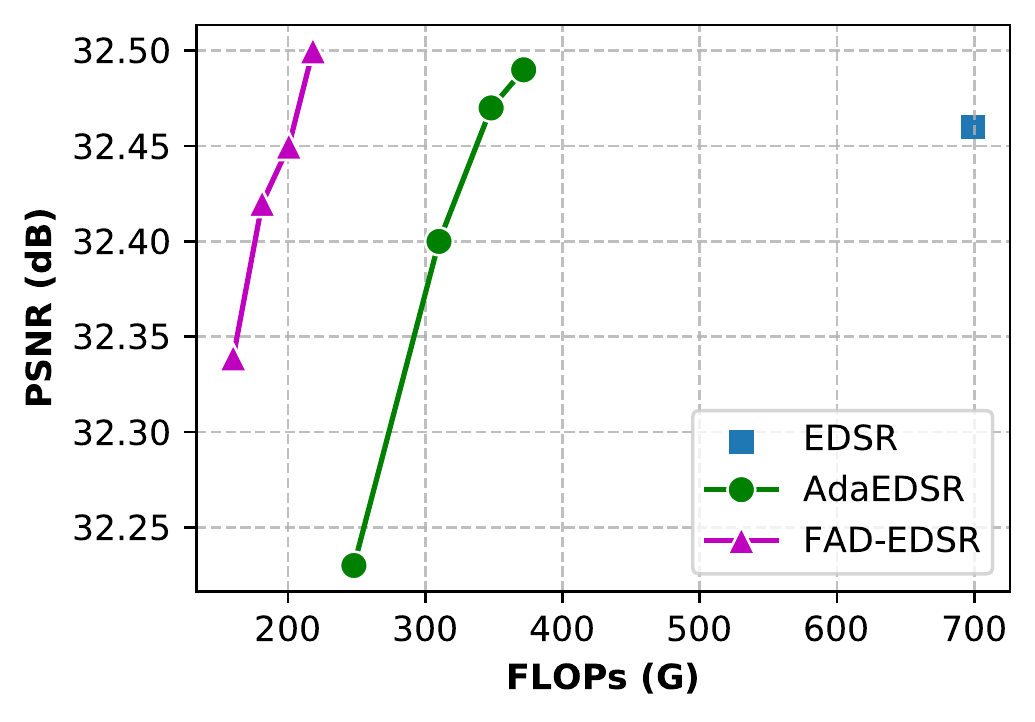}
			\vspace{-0.6cm}   
			\caption{$\times 4$ scale.}
		\end{subfigure}
		\vspace{-0.3cm}
		\caption{Comparison to state-of-the-art methods in terms of PSNR and FLOPs. The benchmark used is Set5.}
		\label{fig:flops_psnr}
		\vspace{-0.6cm}
	\end{center}
\end{figure*}

\section{Experiment}
\label{experiment}
\subsection{Implementation Details}
\paragraph{Training}
We apply the proposed frequency-aware dynamic mechanisms on EDSR~\cite{lim2017enhanced} and RCAN~\cite{zhang2018image} models, which are called FAD-EDSR and FAD-RCAN, respectively. The DIV2K dataset \cite{timofte2017ntire} is adopted to train the dynamic networks. DIV2K includes 800 training images and 100 validation images. The low-resolution (LR) images are generated with bicubic degradation algorithm. The input patch size and data augmentation follow literature~\cite{lim2017enhanced}. The training lasts $800,000$ iterations and is optimized by ADAM algorithm \cite{kingma2014adam} with $\beta_1=0.9$ and $\beta_2=0.999$. The initial learning rate is set to $0.01$ for the mask predictor parameters,  $0.0001$ for other parameters (\ie body parameters) and decays to half every $200,000$ iterations. We set $\alpha$ as $0.4$ and initialize $\beta$ as $0.0001$. We apply a linear annealing strategy on $\beta$ which is reduced to $0$ after $70\%$ iterations. 

\vspace{-0.3cm}
\paragraph{Evaluation}
We evaluate our method with PSNR and SSIM metrics on four benchmarks: Set5~\cite{bevilacqua2012low}, Set14~\cite{zeyde2010single}, B100~\cite{martin2001database} and Urban100~\cite{huang2015single}. The PSNR and SSIM are calculated on the Y channel (\ie the luminance channel) of YCbCr color space. The same amount of pixels as $scale$ from the image boundary are ignored. The computational cost is evaluated by FLOPs (Floating Point of Operations).

\subsection{Comparison with State-of-the-arts}
To evaluate the effectiveness and efficiency of the proposed frequency-aware dynamic mechanism, we compare our experimental results with that of deep adaptive approach \cite{liu2020deep} (abbreviated as Ada-) on prominent EDSR~\cite{lim2017enhanced} and RCAN~\cite{zhang2018image} models. As shown in Table~\ref{table:sota}, compared with original SR model, our method reduces almost 50\% of computations (\ie FLOPs) while keeps and even increases the super-resolution accuracy (\ie PSNR and SSIM). For instance, the FLOPs of EDSR on Set5 $\times 2$ are reduced by $69.6\%$ while PSNR is improved by $0.11dB$. Compared to deep adaptive approach (\ie AdaEDSR and AdaRCAN), our method reduces more computation and gets better SR accuracy. Taking the experiment on B100 with $\times 4$ scale as example, the FLOPs of AdaEDSR are reduced by $22.4\%$ and PSNR is similar to EDSR. In contrast, the FLOPs of our model are reduced more, \ie $53.5\%$, while PSNR is increased by $0.02dB$. An interesting phenomenon is that AdaEDSR meets a serious performance degradation in large-scale upscaling (\ie $\times 3$ and $\times 4$) whose optimizations are harder than $\times 2$ upscaling since the larger-scale downscaling causes more high frequency signals lost. However, our methods still keeps excellent performance because of the end-to-end optimization and coordinated branches. We also evaluated the latency of the proposed frequency-aware dynamic models on CPU platform with different number of cores. The experimental results in Table~\ref{LatencyTable} show that the proposed method reduces the latency of model significantly. 

\begin{figure*}
	\begin{center}
		\includegraphics[width=1\textwidth]{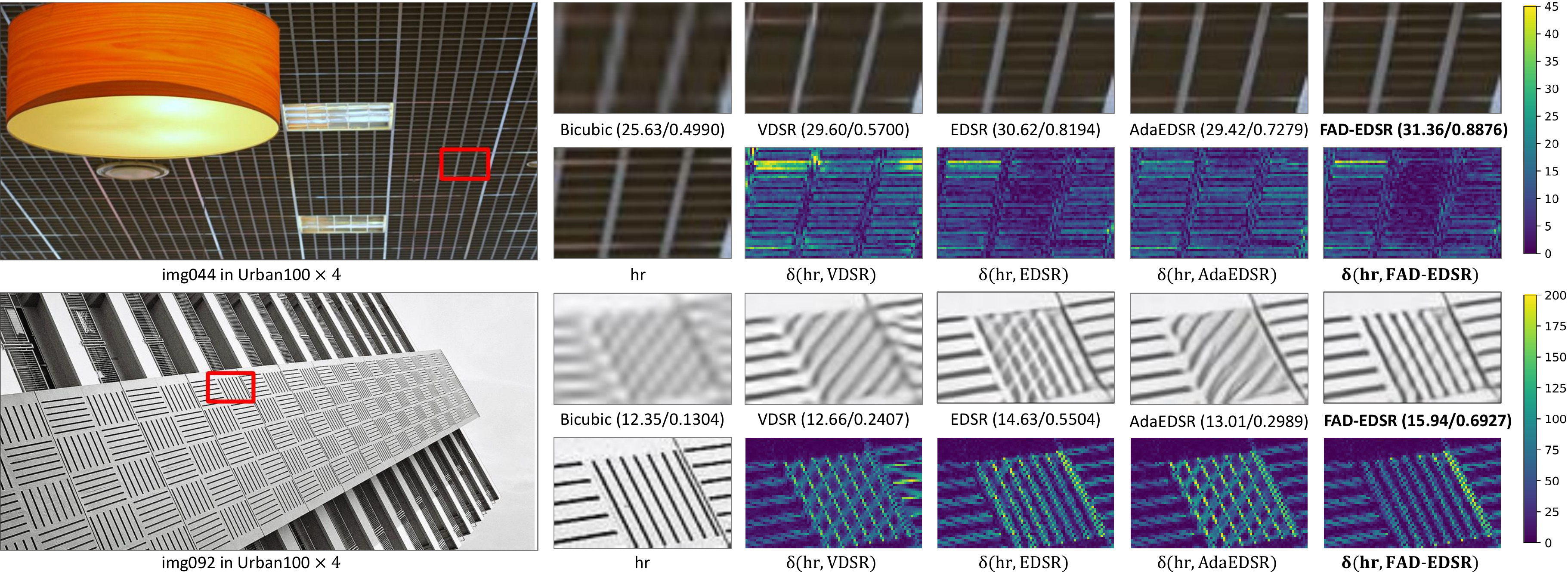}
		\vspace{-0.7cm}
		\caption{Visual Comparison with state-of-the-art methods. $\delta$ is the residual map of the two given images.}
		\label{fig:sr_result}
	\end{center}
	\vspace{-0.6cm}
\end{figure*}

In addition to the quantitative results, we visual the super-resolution resulting images in Fig.~\ref{fig:sr_result} including two common scenes, \ie, blurring and artifacts caused by serious downscaling (the first and second row respectively). The proposed method achieves very close even better visual quality compared with original model. The performance of our method outperforms that of deep adaptive mechanism significantly. We use $\alpha$ to control the computational cost as shown in Equation~\ref{equa:sparsity_loss}. Multiple models with different computations are obtained by adjusting the hyper-parameter and the relation between PSNR and FLOPs is shown in Fig.~\ref{fig:flops_psnr}. The upper left points have better performance. One can see that the proposed method outperforms its counterparts.

\begin{table}[t]
\vspace{-2mm}
\small
\begin{center}
\caption{Latency of various models executed on CPU.}
\vspace{-3mm}
\label{LatencyTable}
\renewcommand{\arraystretch}{1.3}
\setlength{\tabcolsep}{1.0mm}{
\begin{tabular}{|p{2.3cm}<{\centering}|p{1.5cm}<{\centering}|p{1.6cm}<{\centering}|p{2.0cm}<{\centering}|}
\hline
Platform & EDSR & FAD-EDSR & Speed-up ratio\\
\hline
\hline
Single core CPU & 39.60s & 25.44s & \textbf{$\times$1.6} \\
\hline
Double core CPU & 20.08s & 13.74s & \textbf{$\times$1.5} \\
\hline
\end{tabular}}
\end{center}
\vspace{-2mm}
\end{table}

\vspace{-0.3cm}
\paragraph{Why did FADN achieve higher accuracy over baselines?}
The proposed FADN possesses more parameters compared with the original baseline. In addition, the branching strategy allows the weights of each branch to adapt to the specific frequency region instead of covering various frequency regions. Finally, cheap operator can achieve better accuracy than heavy operator on low-frequency regions. Hence, our method achieves better performance than the SR baselines.

\begin{table}[t]
\vspace{-0.2cm}
		\small
		\caption{Comparison of various frequency masks on EDSR-baseline~\cite{lim2017enhanced}. PSNR is reported and scale factor is $\times 2$.}
		\vspace{-0.1cm}
		\begin{tabular}{|p{3.1cm}<{\centering}|p{0.65cm}<{\centering}|p{0.65cm}<{\centering}|p{0.65cm}<{\centering}| p{1.1cm}<{\centering}|}
			\hline
			Methods & Set5 & Set14 & B100 & Urban100 \\
			\hline
			Random for blocks & 37.79 & 33.33  & 32.06  & 31.47  \\
			\hline
			Random for images & 37.67  & 33.19  & 31.95 & 31.17 \\
			\hline
			DCT mask & 37.82 & 33.40 & 32.07 & 31.72 \\
			\hline
			Learnable w/o guidance & 37.86 & 33.43 &  32.11  & 31.81  \\
			\hline
			Learnable w/ guidance  & {\bf 37.92}  & {\bf 33.49} & {\bf 32.13}  & {\bf 31.90}  \\
			\hline
		\end{tabular}
		\label{table:frequency_map}
		\vspace{-0.3cm}
\end{table}

More prominent SISR models are evaluated in the {\color{red} Supplementary Material}. It shows that the proposed method reduces FLOPs significantly while preserves the performance.

\subsection{Ablation Study}
\label{ablation study}
In ablation study, we apply our method on EDSR baseline~\cite{lim2017enhanced} which includes 16 residual blocks of 64 channels. All models are trained for $300,000$ iterations and the learning rate decays to half every $100,000$ iterations. Note that We have adjusted the hyper-parameter $\alpha$ and guaranteed the computations among each experiment are comparable. 

\vspace{-0.4cm}
\paragraph{Learnable Frequency Mask}

\begin{figure}[!t]
	\captionsetup{font={footnotesize}}
	\begin{minipage}[t]{0.48\columnwidth}
		\includegraphics[width=\textwidth]{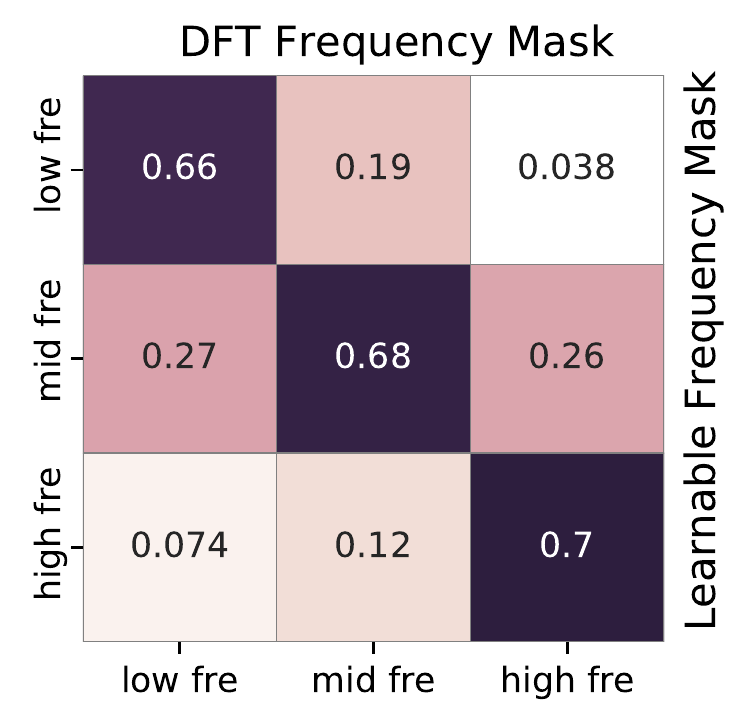}
	     \vspace{-0.7cm}
		\caption{Confusion Matrix between DCT and learnable frequency mask.}
		\label{fig:Confusion Matrix}
	\end{minipage} 
	\hspace{0.1cm}
	\begin{minipage}[t]{0.48\columnwidth}
		\includegraphics[width=\textwidth]{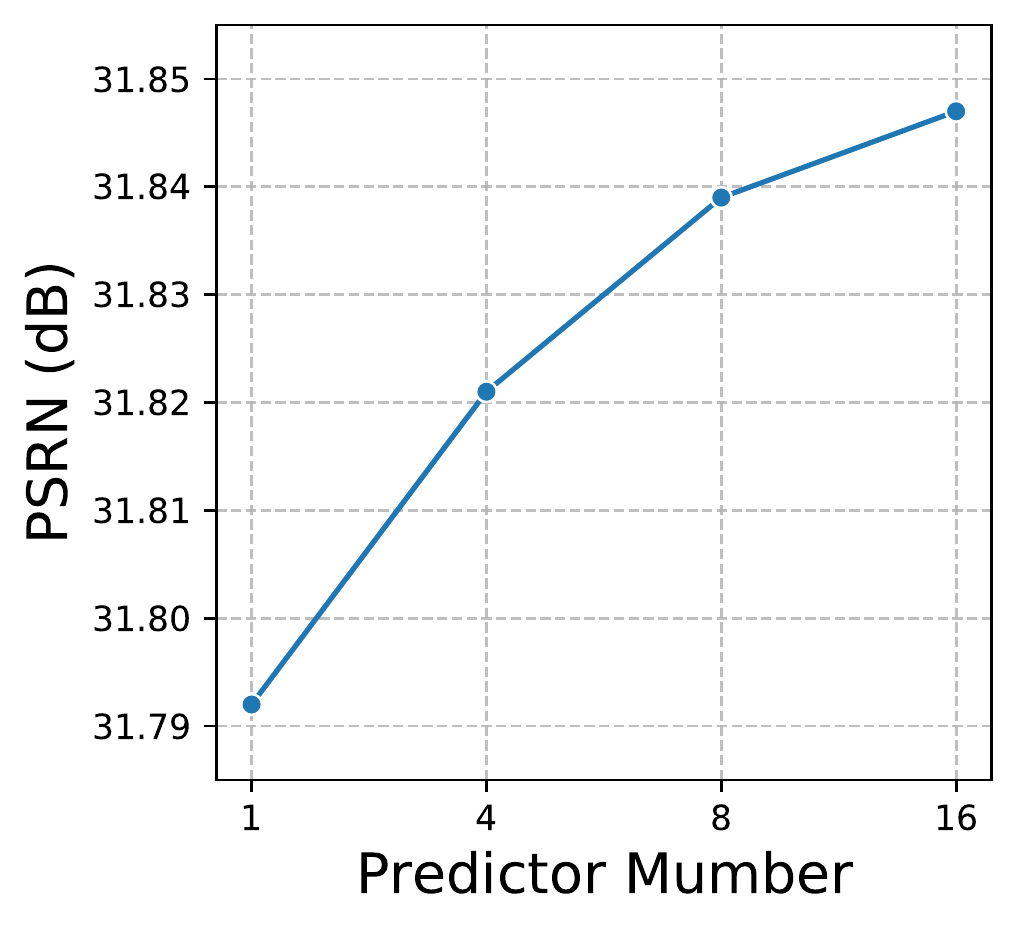}
	      \vspace{-0.7cm}
		\caption{Comparison of different predictor number. PSNR in Urban $\times 2$ is reported.}
		\label{fig:predictor_number}
	\end{minipage}
	\vspace{-0.3cm}
\end{figure}

\begin{figure*}[t]
	\begin{center}
		\includegraphics[width=1\textwidth]{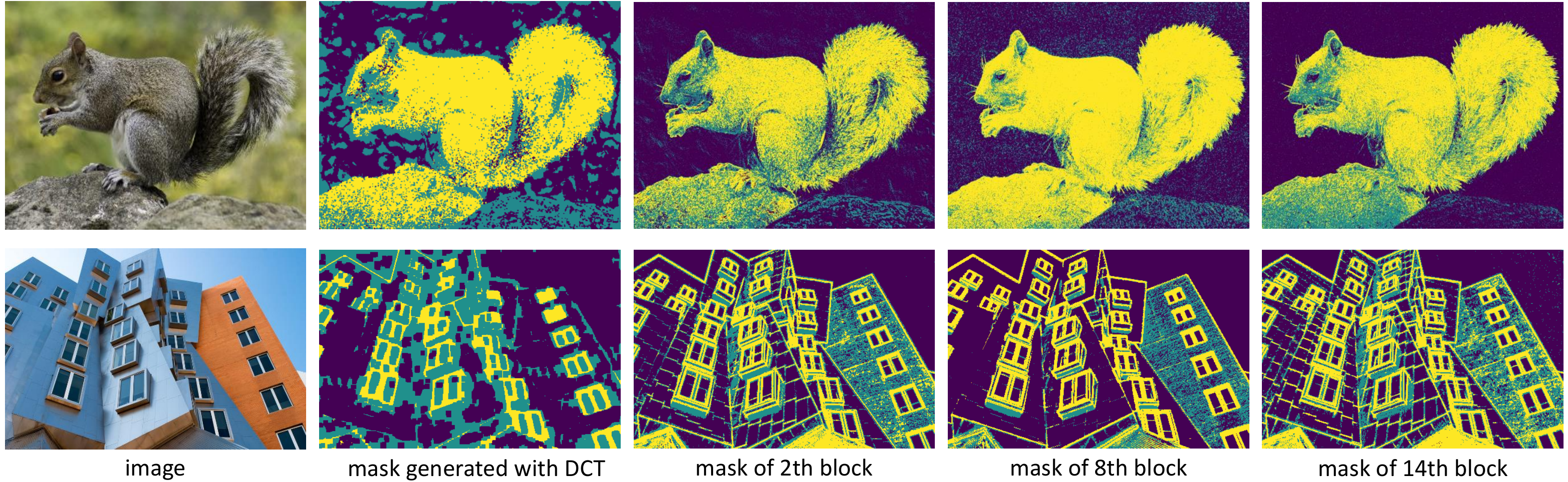}
		\vspace{-0.7cm} 
		\caption{Comparison between frequency masks generated with DCT and our learnable frequency masks in different blocks. Yellow, Blue and Purple represent High, Medium and Low frequency respectively.}
		\label{fig:frequency_map}
	\end{center}
	\vspace{-0.7cm}
\end{figure*}

The frequency masks are learned by an end-to-end manner in our proposed methods. To demonstrate the effectiveness of learnable frequency masks, we exploit different strategies for mask generation, including {\it random masks for each block}, {\it random masks for each image (\ie, sharing a fixed random mask among blocks)}, {\it fixed frequency masks generated by DCT}, {\it learnable frequency masks without DCT guidance} and {\it learnable frequency masks with DCT guidance}. Note that we are more concerned about the performance on Urban100 since its image quantity and quality are higher. 
As shown in Table.~\ref{table:frequency_map}, the performance using random masks is terrible because the model cannot identify the frequency information and assign pixels to the reasonable branches. A fixed random mask is worse than multiple random masks among blocks. The DCT frequency masks outperform random masks because the former provide appropriate indication that assigns pixels to the proper branches. However, the DCT frequency masks are dependent on human-crafted thresholds and sensitive to noise. Besides, the specific thresholds cannot adapt to all images. The learnable frequency masks avoid the problem by end-to-end optimization. Without the guidance of DCT prior knowledge, learnable frequency masks outperform DCT frequency masks. With the help of guidance, models can converge faster and achieve better performance. 

Apart from the quantitative performance comparison, we compare the learnable frequency masks with DCT frequency masks. As shown in Fig.~\ref{fig:frequency_map}, DCT frequency masks indicate the frequency distribution but are sensitive to noises. Our learnable frequency masks are similar to the DCT frequency masks except for minority regions where the predictor adjusts the mask according to specific inputs for more accurate indication. Besides, the learnable frequency masks are little different among each block since it's not a wise choice to use a fixed frequency mask for all blocks. Generally, the lower frequency pixels tend to use the branches requiring less computations. The result implies the proposed FADN model can identify the frequency strength and assign the pixels to branches of proper computations according to the frequency strength. The confusion matrix between DCT frequency masks and learnable frequency masks is shown in Fig.~\ref{fig:Confusion Matrix} which also proves that the mask predictor can learn the frequency distribution and adjust the partition thresholds automatically.

\vspace{-0.3cm}
\paragraph{Mask Predictor Number}
By default, we use a separate mask predictor for each frequency-aware dynamic block. Besides, the predictor can also be shared among multiple blocks. We visualize the influence of predictor number in Figure~\ref{fig:predictor_number}. In partitularly, the model learns one frequency mask for the whole network when the predictor number is 1 and learns an independent mask for each block when the number is 16. One can see that a single predictor performs worse than multiple predictors, because the optimal frequency partition is not invariant among blocks. It is another important reason why we can not use the fixed frequency masks generated with DCT directly. Overall, as the predictor number increases, the adjustment ability of FADN is more powerful. The performance therefore becomes better although the improvement grows more slowly.

\vspace{-0.3cm}
\paragraph{Branch Number}
\label{setction:branchnumber}
We use 3-branch FADN in above experiments but the branch number and branch architectures are optional. We test the influence of different branch number. As the branch number increases, each branch only needs to recover signals in a smaller frequency range. The results are shown in Table.~\ref{table:branch_number}. Note that we have guaranteed similar computations among different settings. The performance of 3-branch is significantly better than 2-branch while 4-branch is slightly better than 3-branch. Considering the storage cost, 3-branch FADN are recommended.

\begin{table}[t]
    \caption{Comparison of different branch number. PSNR is reported and scale factor is $\times 2$.}
	\vspace{-0.7cm}
    \begin{center}
    \small
    \begin{tabular}[width=1\columnwidth]{|b{1.8cm}<{\centering}|b{1.1cm}<{\centering}| b{1.1cm}<{\centering}|b{1.1cm}<{\centering}| b{1.1cm}<{\centering}|}
    \hline
    Branch Num. & Set5 & Set14 & B100 & Urban100 \\
    \hline
    2 &  37.84 & 33.44 & 32.09 & 31.72 \\
    \hline
    3 & {\bf 37.90} & 33.47 & {\bf 32.13}  & 31.85 \\
    \hline
    4 & {\bf 37.90} & {\bf 33.48} & 32.12 & {\bf 31.88}  \\
    \hline
\end{tabular}
\label{table:branch_number}
\vspace{-0.8cm}
\end{center}
\end{table}

\section{Conclusion}
\label{conclusion}
In this paper, a novel frequency-aware dynamic network (FADN) is proposed for efficient single image super resolution, which assigns cheap operations to low-frequency regions and vice visa. To this end, a predictor is introduced to divide the input feature into multiple components based on DCT domain. The predictor is learned under the supervision of hand-crafted frequency-domain masks of images in the training set and the reconstruction loss of SISR, simultaneously. The overall computational complexity will be significantly reduced by the optimized allocation. In addition, the frequency-aware dynamic mechanism can be conveniently employed on various SISR architectures. Experimental results indicate that FADN can effectively reduce approximate half of FLOPs in multiple benchmark databases while maintaining the performance of the original network. 

{\small
\bibliographystyle{ieee_fullname}
\bibliography{ref}
}

\end{document}